\documentclass[twocolumn,superscriptaddress,notitlepage,nofootinbib]{revtex4-2}
\usepackage{graphicx}
\usepackage{amsmath,amssymb,setspace}
\usepackage{newtxtext}
\usepackage{newtxmath}
\usepackage{ulem}
\usepackage{multirow}
\usepackage[x11names]{xcolor}
\usepackage[unicode,colorlinks,citecolor=Blue3,linkcolor=Blue3,bookmarks=true]{hyperref}

\newcommand{\eg}{{\it e.g.\,}}

\newcommand{\mean}[1]{\langle{#1}\rangle}
\renewcommand{\Re}{\mathop{\rm Re}\nolimits}
\renewcommand{\Im}{\mathop{\rm Im}\nolimits}
\newcommand{\fulld}[2]{\dfrac{d#1}{d#2}}
\newcommand{\sh}{\sinh}
\newcommand{\ch}{\cosh}

\begin{document}

\title{Application of optical squeezing to microresonator based optical sensors}

\date{\today}

\author{Dariya Salykina}%
\email{koil257@mail.ru}
\affiliation{Faculty of Physics, M.V. Lomonosov Moscow State University, Leninskie Gory 1, Moscow 119991, Russia}
\affiliation{Russian Quantum Center, Skolkovo IC, Bolshoy Bulvar 30, bld.\ 1, Moscow, 121205, Russia}

\author{ Daniil Shakhbaziants}
\affiliation{Russian Quantum Center, Skolkovo IC, Bolshoy Bulvar 30, bld.\ 1, Moscow, 121205, Russia}
\affiliation{Moscow Institute of Physics and Technology, 9 Institutsky Lane, Dolgoprudny, 141701, Russia}

\author{Igor Bilenko}
\affiliation{Faculty of Physics, M.V. Lomonosov Moscow State University, Leninskie Gory 1, Moscow 119991, Russia}
\affiliation{Russian Quantum Center, Skolkovo IC, Bolshoy Bulvar 30, bld.\ 1, Moscow, 121205, Russia}

\author{Farid Ya.\ Khalili}
\email{farit.khalili@gmail.com}
\affiliation{Russian Quantum Center, Skolkovo IC, Bolshoy Bulvar 30, bld.\ 1, Moscow, 121205, Russia}


\begin{abstract}
  High-Q optical microresonators combine low losses and high optical energy concentration in a small effective mode volume, making them an attractive platform for optical sensors. While light is confined in the microresonator by total internal reflection, a portion of the optical field, known as the evanescent field, extends outside. This makes the mode's resonant frequency sensitive to changes in the surrounding environment.

  In this work, we explore the quantum sensitivity limits of this type of sensors. We show that using the intracavity squeezing of the light in the microresonator, it is possible to suppress the influence of the optical losses and cancel the undesirable self phase modulation effect, originating from the cubic non-linearity of the microresonators media. As a result, the sensitivity surpassing the shot noise limit can be achieved. An additional sensitivity gain can be obtained by preparing the input light in a squeezed quantum state.
\end{abstract}

\maketitle

\section{Introduction}

High-$Q$ optical microresonators \cite{braginsky1989quality, strekalov2016nonlinear} have unique properties that make them an attractive tool for optical sensors. The combination of low losses and high optical energy in a small effective mode volume allows for high sensitivity to various external influences. Progress in the fabrication of microresonators has led to the development of integrated technologies, thanks to which a microresonator sensor can be manufactured on a small chip and still have high enough quality factor and nonlinearity values \cite{Shi:23, Zhang:17, Zhang:21}.

While probe light propagates within the microresonator due to almost total internal reflection, some of the optical field escapes from the microresonator, forming an evanescent field \cite{Gorodetsky1994HighQOW, Righini2011WhisperingGM}. Any change in the environment within the area of this evanescent field leads to a shift in the resonant frequency of the microresonator, which can be detected by measuring the output phase of the probe field. Such sensor could be relevant for biosensors and  analysis of chemical reactions, up to distinguishing single molecules \cite{zhang2018optical, granizo2025functionalized}. Theoretical \cite{yu2021whispering, arnold2013taking} and experimental \cite{vollmer2008single, li2018quantum, yue2021label} studies have been conducted to demonstrate the practical effectiveness of a sensor based on microresonators.

It is well known that the sensitivity of optical phase measurements is limited by quantum effects and specifically by the shot noise limit (SNL) \cite{caves1981quantum}. In order to overcome this limit, non-classical states of the probe light, such as squeezed states, have to be used \cite{caves1981quantum, balybin2023improving}. This approach promises the way to keep the field strengths well below a photodamage threshold of a phase shifting sample  while maintaining the required sensitivity and measurement speed \cite{casacio2021quantum, zossimova2024whispering, butt2025beyond}.

Unfortunately, the gain provided by the squeezed light is sensitive to the optical losses, in particular, to the photodetection inefficiency. This problem can be alleviated by using an additional anti-squeezing (parametric amplification) of the output beam, as it was proposed in Ref.\,\cite{caves1981quantum} and demonstrated experimentally in Ref.\,\cite{frascella2021overcoming}. Note that the anti-squeezer should be placed {\rm before} the loss source. Therefore, in order to mitigate the losses created by the coupling of the microresonators with the output path (an important source of losses in the microresonators), parametric intracavity excitation \cite{peano2015intracavity, korobko2023mitigating} should be used. It can be implemented by means of three- or four-wave mixing processes. In the first case, the quadratic  nonlinearity is required, and in the second case, the cubic one; in both cases, the net result is the same. In this work, we consider (i) input mode in a squeezed state to improve the sensitivity and (ii) intracavity squeezing to reduce external noise and impact of negative nonlinear effects.

Typically, the cubic nonlinearity is present in the high-quality microresonator media, and the high concentration of the optical energy makes the nonlinear effects well pronounced. In Ref.\,\cite{salykina2025intracavity}, the use of $\chi^{(3)}$ nonlinearity was discussed in the context of the QND measurements of the optical power, see Refs.\,\cite{77a1eBrKhVo, 81a1eBrVy, Milburn_PRA_28_2065_1983, Roch_APB_55_291_1992, 96a1BrKh, Drummondquantum1994, Balybin2022}. In that case, two beams, propagating through the microresonator, the signal and probe ones, interact by means of the cross-phase modulation effect, originating from the cubic nonlinearity. Subsequent measurement of the probe beam phase allows to extract the value of the signal mode power without (in the ideal lossless case) changing it. On the other hand, the same cubic nonlinearity also creates the self-phase modulation effect (SPM) of the probe beam \cite{yuen1979generation} which could degrade the sensitivity of the microresonator-based sensors \cite{Drummondquantum1994}. It was shown in Ref.\,\cite{salykina2025intracavity}, that the intracavity parametric interaction allows to mitigate this effect.

In this paper, we analyze the quantum-imposed sensitivity limits for a general microresonator-based optical sensor of small variations of eigen frequency, that could be imposed by any external agent, see Fig.\ref{sensor}. Similarly to Ref.\,\cite{salykina2025intracavity}, we assume that the probe beam is prepared in the squeezed quantum state and additionally is parametrically (anti-)squeezed inside the cavity. We extend the analysis of that paper by performing the rigorous optimization of input squeezing and the output homodyne detection angles, as well as of the intracavity (anti-)squeezing parameters (in the previous paper, they were chosen for heuristic reasons).

The paper is organized as follows. In the next section, we calculate the quantum noise of the proposed sensor. Then in Sec.\,\ref{sec:opt} we optimize the spectral density of this noise. Finally, in Sec.\,\ref{sec:conclusion} we provide estimates for achievable sensitivity and summarize the obtained results. 

\section{Quantum noise}\label{sec:calcs}

\begin{figure}
\includegraphics[width=0.5\textwidth]{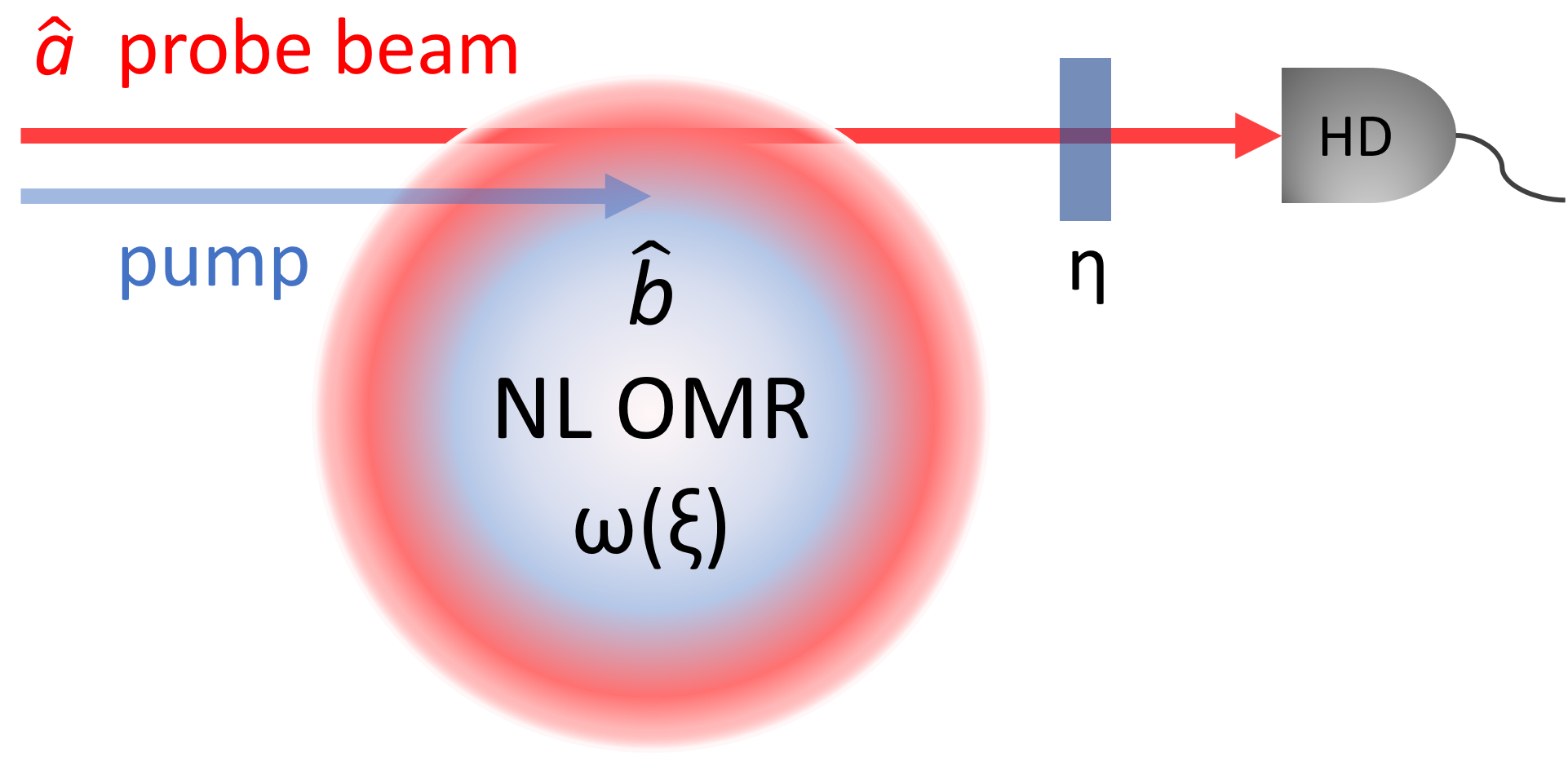}
\caption{\label{sensor} Schematic illustration of optical sensor based on microresonator. NL OMR -- nonlinear optical microresonator, HD -- homodyne detector (the corresponding reference beam is not shown for simplicity), $\hat a$ -- input mode in a squeezed coherent state, $\eta$ is the quantum efficiency of the output path, including the photodetection quantum efficiency. The additional ``pump'' provides the intracavity parametric interaction}.
\end{figure}

Let us consider an optical mode of the microresonator described by annihilation operator $\hat{b}$, see Fig.\,\ref{sensor}. We suppose that its eigen frequency fluctuates slightly around the mean value $\omega_o$ due to some small external time-dependent factor $\xi(t)$:
\begin{equation}
  \omega(t) = \omega_o + \xi(t) \,,\quad |\xi|\ll\omega_o \,.
\end{equation}
Our goal is to measure this factor.

We suppose that this mode can be parametrically excited using three-wave or four-wave mixing processes. We assume also that in any case the third-order nonlinearity is present in the scheme, leading to the SPM effect.

The corresponding Hamiltonian can be presented as follows:
\begin{multline}\label{hamilt}
  \frac{\hat H}{\hbar}
  = (\omega_o + \xi)\hat{b}^{\dagger}\hat{b}
    - \frac{\gamma}{2}\hat{b}^{\dagger 2}\hat{b}^2
    + \frac{ik}{2}(
          \hat{b}^{\dagger 2}e^{-2i\omega't - i\phi} - \hat{b}^2e^{2i\omega't + i\phi}
        ) \\
     + \text{coupling terms} \,,
\end{multline}
$\gamma$ is the self-phase modulation factor (see {\it e.g.} Eq.\,(61) of Ref.\,\cite{balybin2020perspectives}), $k$ is the intracavity parametric interaction factor, $2\omega'$ is the frequency of the parametric interaction and $\phi$ is its phase. The ``coupling terms'' describe the intracavity losses and the coupling of the probe mode with the input and output paths; see details in \eg Sec.\,III\,B of the review  \cite{aspelmeyer2014cavity}. We model the output losses by the fictional beamsplitter with the power transmissivity $\eta$ \cite{leonhardt1995measuring, yuen1978quantum}.

The corresponding Heisenberg-Langevin equations of motion are the following:
\begin{multline}\label{eqs_raw_b}
  \fulld{\hat{b}}{t} + \kappa\hat{b} + i(\omega_o + \xi)\hat{b}
  - i\gamma\hat{b}^\dag\hat{b}^2 - k\hat{b}^\dag e^{-2i\omega't-i\phi} \\
  = \sqrt{2\kappa'}\hat{a} + \sqrt{2\kappa''}\hat{v} \,,
\end{multline}
\begin{equation}\label{eqs_raw_d}
  \hat{d} = \sqrt{\eta}(\sqrt{2\kappa'}\hat{b} - \hat{a} + \epsilon\hat u) \,,
\end{equation}
where
\begin{gather}
  \epsilon = \sqrt{\frac{1-\eta}{\eta}} \,, \label{eps}\\
  \kappa = \kappa' + \kappa''
\end{gather}
is the half-bandwidth of the mode resonance, $\kappa'$, $\kappa''$ are its parts originating, respectively, from the coupling with the input/output field and from the internal optical losses in the mode, $\hat{a}$, $\hat{d}$ are the annihilation operators of, respectively, the input and effective (with account for the losses) output running waves,  $1-\eta$ is the optical losses factor of the output channel, including the photodetection inefficiency, and $\hat{u}$,  $\hat{v}$ are the annihilation operators of the corresponding noises. The running wave operators $\hat{a}$, $\hat{b}$, $\hat{d}$, $\hat{u}$, $\hat{v}$ are normalized by the commutation relations that, for all of them, have the following form:
\begin{equation}
  [\hat{a}(t),\hat{a}^\dag(t')] = \delta(t-t') \,
\end{equation}
We assume that the input beam $a$ is prepared in the squeezed coherent state, while the noise fields $u$, $v$ are in the ground state.

Now, let us consider the frame rotating at some frequency $\omega'$:
\begin{equation}
  \hat{a} \to \hat{a}e^{-i\omega't} \,
\end{equation}
and similarly, for  $\hat{b}$, $\hat{d}$, $\hat{u}$, $\hat{v}$. In this frame, Eq.\,\eqref{eqs_raw_b} takes the following form:
\begin{equation}\label{eqs_RWA0}
  \fulld{\hat{b}}{t} + \kappa\hat{b} + i(\omega_o + \xi - \omega')\hat{b}
      - i\gamma\hat{b}^\dag\hat{b}^2 - k\hat{b}^\dag e^{-i\phi}
    = \sqrt{2\kappa'}\hat{a} + \sqrt{2\kappa''}\hat{v} \,, \\
\end{equation}
equation \eqref{eqs_raw_d} for $\hat{d}$ remains unchanged. Let us also explicitly detach the mean values from the quantum fluctuations:
\begin{equation}
  \hat{a}(t) \to \alpha + \hat{a}(t) \,, \quad \hat{b}(t) \to \beta + \hat{b}(t)\,, \quad
    \hat{d}(t) \to \delta + \hat{d}(t) \,,
\end{equation}
where $\alpha$, $\beta$, $\delta$ are the strong classical amplitudes and $\hat{a}$, $\hat{b}$, $\hat{d}$ are now quantum fluctuations of the respective modes with $\mean{\hat{a}}=0$. We assume that $\alpha$, $\beta$, $\delta$ do not depend on time. Without limiting the generality, we also assume that $\Im\beta=0$. Keeping then in Eqs.\,\eqref{eqs_RWA0} only linear in quantum fluctuations and in the signal $\xi$ terms and assuming that
\begin{equation}
  \omega' = \omega_o-\gamma N \,,
\end{equation}
where
\begin{equation}
  N = \beta^2
\end{equation}
is the mean photon number in the cavity, we come to the following linear equation of motion:
\begin{equation}\label{eqs_RWA}
  \fulld{\hat{b}}{t} + \kappa\hat{b} - i\gamma N(\hat{b} + \hat{b}^\dag)
    - k\hat{b}^\dag e^{-i\phi}
  = -i\beta\xi + \sqrt{2\kappa'}\hat{a} + \sqrt{2\kappa''}\hat{v} \,. \\
\end{equation}
Equation for $\hat{d}$ again retains the form \eqref{eqs_raw_d}.

Now we introduce the cosine and sine two-photon quadrature amplitudes (see Refs.\,\cite{caves1985new, schumaker1985new}):
\begin{equation}
  \hat{a}_c(\Omega) = \frac{\hat{a}(\Omega) + \hat{a}^\dag(-\Omega)}{\sqrt{2}} \,, \quad
  \hat{a}_s(\Omega) = \frac{\hat{a}(\Omega) - \hat{a}^\dag(-\Omega)}{i\sqrt{2}} \,,
\end{equation}
and similarly for all other optical fields. Here $\Omega$ is a detuning from resonant frequency $\omega'$. In these notations, using Fourier representation, Eqs.\,\eqref{eqs_RWA} and \eqref{eqs_raw_d} can be presented as follows:
\begin{subequations}\label{b_cs}
  \begin{gather}
    (\ell - k_c)\hat{b}_c + k_s\hat{b}_s
      = \sqrt{2\kappa'}\hat{a}_c + \sqrt{2\kappa''}\hat{v}_c \,,\label{eqs_quad_s_raw0} \\
    (\ell + k_c)\hat{b}_s + k_\gamma\hat{b}_c
      = -\sqrt{2}\beta\xi + \sqrt{2\kappa'}\hat{a}_s + \sqrt{2\kappa''}\hat{v}_s \,\label{eqs_quad_s_raw},
  \end{gather}
\end{subequations}
\begin{equation}\label{d_cs}
  \hat{d}_{c,s}
  = \sqrt{\eta}(\sqrt{2\kappa'}\hat{b}_{c,s} - \hat{a}_{c,s} + \epsilon\hat u_{c,s}) \,,
\end{equation}
where
\begin{gather}
  \ell = \kappa - i\Omega \,, \\
  k_c = k\cos\phi \,, \quad k_s = k\sin\phi \,, \\
  k_\gamma = k_s - 2\gamma N \,.
\end{gather}

The second of these equations contains the signal factor $\propto\xi$ in its r.h.s. Due to this reason, measurement of the corresponding quadrature $\hat{b}_s$ was considered in Ref.\,\cite{salykina2025intracavity}. In order to mitigate the influence of the SPM, it was proposed in that work to set
\begin{equation}\label{k_s_opt}
  k_s=2\gamma N \,.
\end{equation}
However, the equations \eqref{b_cs} remain interconnected due to the term $k_s\hat b_s$ in the l.h.s. of Eq.\,\eqref{eqs_quad_s_raw0}, and the cosine quadrature contains information on signal $\xi$.

In this paper, we initially assume that some optimal quadrature contains maximal information on signal $\xi$:
\begin{equation}\label{d_zeta_raw}
  \hat{d}_\zeta = \hat{d}_c\cos\zeta + \hat{d}_s\sin\zeta \,,
\end{equation}
and therefore is measured by the homodyne detector, where $\zeta$ is the homodyne angle. The explicit form of $\hat{d}_\zeta$ is quite involving. It is calculated in Appendix \ref{app:d_zeta}, where it is presented in the following convenient form, see Eqs.\,\eqref{d_zeta_app}-\eqref{A_11},\,\eqref{eps}:
\begin{equation}\label{d_zeta}
  \hat{d}_\zeta = K\times(\xi + \xi_{\rm fl}) \,,
\end{equation}
where $K$ is the common normalization prefactor that does not affect the sensitivity and $\hat{\xi_{\rm fl}}$ is the normalized noise, reduced to the meter input.

\section{Optimization of the noise}\label{sec:opt}

It follows from Eq.\,\eqref{d_zeta} that the sensitivity of the considered scheme is defined by the spectral density $S_{\xi}$ of noise $\xi_{\text{fl}}$, which is reduced to the input, see Eq.\,(\ref{S_xi}). We minimize it in four parameters: squeeze angle $\theta$, homodyne angle $\zeta$ and the internal squeezing parameters $k_c$, $k_s$. Evidently, the optimal values of these parameters depend on the frequency $\Omega$ at which the optimization is performed (we do not consider the sophisticated frequency dependent squeezing and homodyne angle technologies of Ref.\,\cite{02a1KiLeMaThVy}). Here we consider optimization at the best sensitivity point of $\Omega=0$ (the resonance case).


We perform the optimization in three steps. First, we optimize $S_\xi$ analytically for the first two parameters $\theta$ and $\zeta$, see Appendix \ref{app:opt_theta_zeta_1}.  As the second step, we numerically minimize the resulting spectral density \eqref{S_W} in $k_s$.
As a result we obtain, that within the precision limit of numerical calculation, the optimal value of $k_s$ is given by Eq.\,\eqref{k_s_opt}, which corresponds to the cancellation of the SPM, see Appendix\,\ref{app:opt_stage2}. Third, we substitute this value of $k_s$ into the equations \eqref{W},\,\eqref{S_W} and, using the {\sf Wolfram Mathematica}, analytically find the optimal value for $k_c$, which is equal to
\begin{equation}\label{k_c_opt}
  k_c = \frac{(\kappa' - \kappa'')e^{-2r} - \epsilon^2\kappa}{e^{-2r} + \epsilon^2}.
\end{equation}
Substitution of this value into Eqs.\,\eqref{opt_zeta}, \eqref{opt_theta} gives the following optimal values of the homodyne and squeeze angles, see Appendix \ref{app:opt_theta_zeta_3}:
\begin{subequations}\label{zeta_theta}
  \begin{gather}
    \cos\zeta = 0 \,, \label{zeta_opt} \\
    \tan2\theta = 0 \,.\label{theta_opt}
  \end{gather}
\end{subequations}
Thus, the sine quadrature should be squeezed at the input and measured by the homodyne detector.

It is interesting to note that these values, obtained by the rigorous optimization, are in agreement with the ones guessed in our previous paper \cite{salykina2025intracavity}, see Eq.\,(26) of that paper.

Note also that the sign of the optimal $k_c$ depends on whether the input noise (the factor $e^{-2r}$) or the noise introduced by the output losses (the factor $\epsilon^2$) dominates. In the first case, $k_c>0$, which corresponds to the squeezing of the measured (sine) quadrature, see Eq.\,\eqref{eqs_quad_s_raw}, as it was first proposed in Ref.\,\cite{peano2015intracavity}. In the second case, $k_c<0$, which corresponds to the anti-squeezing of this quadrature \cite{korobko2023mitigating, salykina2025intracavity}.

Substituting the values \eqref{k_s_opt} and \eqref{zeta_opt} into Eq.\,\eqref{d_zeta_app}-\eqref{A_11}, we obtain:
\begin{equation}
  \hat{\xi}_{\rm fl} = \frac{1}{-2\sqrt{\kappa'}\beta}\bigl[
    (\kappa'-\kappa''-k_c+i\Omega)\hat{a}_s + 2\sqrt{\kappa'\kappa''}\hat{v}_s
    + \epsilon(\kappa+k_c-i\Omega)\hat{u}_s
  \bigr] .
\end{equation}
Finally, taking into account Eqs.\,\eqref{S_11},\,\eqref{k_c_opt}, and \eqref{theta_opt}, we obtain the optimized spectral density of $\hat{\xi}_{\rm fl}$:
\begin{equation}\label{S_3}
  S_\xi(\Omega) = \frac{1}{2N}\biggl(
      \frac{e^{-2r} + \epsilon^2}{4\kappa'}\Omega^2
      + \frac{\epsilon^2\kappa'}{1 + \epsilon^2e^{2r}} + \kappa''
    \biggr) .
\end{equation}

\section{Discussion}\label{sec:conclusion}

In order to demonstrate advantages, provided by both input and internal squeezing, we compare the following three characteristic scenarios.
\begin{enumerate}
  \item No squeezing at all:
    \begin{equation}\label{no_sqz}
      r=0\,, \quad k=0 \,.
    \end{equation}
    It is shown in Appendix \ref{app:no_sqz}, that in this case, the spectral density of $\hat{\xi}_{\rm fl}$ is equal to
    \begin{equation}\label{S_0}
      S_\xi(\Omega) = \frac{1}{2N}\biggl[
          \frac{\kappa^2+\Omega^2}{4\eta\kappa'}
          + \frac{1}{\kappa^2+\Omega^2}
              \biggl(\frac{\eta\kappa'\Omega^4}{\kappa^4} + \kappa - \eta\kappa'\biggr)
              k_\gamma^2
      \biggr] .
    \end{equation}
    Note that the second term in the square brackets, originating from the uncompensated SPM, remains. Parameter $k_{\gamma}$ depends on $N$, therefore, with the growth of photon number, impact of this term will increase, worsening the sensitivity of the scheme.
  \item Internal squeezing only:
    \begin{equation}
      r=0\,, \quad k\ne0 \,.
    \end{equation}
    In this case, setting in Eq.\,\eqref{S_3} $r=0$, we obtain:
    \begin{equation}\label{S_1}
      S_\xi(\Omega) = \frac{1}{2N}
        \biggl[\frac{\Omega^2}{4\eta\kappa'} + (1-\eta)\kappa' + \kappa''\biggr] .
    \end{equation}
    Here, the SPM term is fully canceled. Moreover, the remaining zero-frequency term is also significantly suppressed.
  \item Both input squeezing and internal squeezing are used:
    \begin{equation}
      r>0\,,\quad k\ne 0 \,.
    \end{equation}
    In this case, the sensitivity is described by Eq.\,\eqref{S_3}, which shows the additional advantage provided by the input squeezing.
\end{enumerate}

One more possible option for improving the sensitivity is the use of anti-squeezing (parametric amplification) in the output path before detection. It is easy to show that in this case, assuming the optimal value of the anti-squeeze angle equal to the homodyne angle $\zeta$, Eqs.\,\eqref{S_0}, \eqref{S_1}, and \eqref{S_3} retain their form, but with the output losses factor suppressed as follows:
\begin{equation}
  \epsilon^2 \to \epsilon^2e^{-2R} \,,
\end{equation}
where $R$ is the anti-squeeze factor.

It is instructive to compare the obtained results with the SNL, that is, the best sensitivity achievable without the squeezing. In order to derive it, we set
\begin{equation}
  \eta = 1 \,, \quad \kappa'' = 0 \,, \quad k_\gamma = 0
\end{equation}
in Eq.\,\eqref{S_0}, obtaining the following idealized spectral density:
\begin{equation}
  S_\xi(\Omega) = \frac{\kappa^2 + \Omega^2}{8N\kappa} \,.
\end{equation}
Then, we assume the value of $\kappa=|\Omega|$ that minimizes this equation. The resulting SNL is equal to:
\begin{equation}\label{S_SNL}
  S_{\rm SNL}(\Omega)  = \frac{|\Omega|}{4N} \,.
\end{equation}

\begin{table}
  \begin{ruledtabular}
    \begin{tabular}{ccc}
      Quantity & Notation & Value \\ \hline
      Wavelength &$\lambda$ & $1.5\mu{\rm m}$ \\
      Q-factor & $Q$ & $10^9$ \\
      Intrinsic bandwidth & $\kappa''$ & $0.6\times10^6\,{\rm s}^{-1}$ \\
      Loaded bandwidth & $\kappa $ & $10\kappa''$ \\
      Nonlinearity & $\gamma$ & $0.08\,{\rm s}^{-1}$ \\
      Quantum efficiency & $\eta$ & $0.9$ \\
      Mean photon number & $N$ & $10^8$ \\
      Squeeze factor & $e^{2r}$ & 30 (15\,dB)
    \end{tabular}
  \end{ruledtabular}
\caption{Parameters values used for the estimates.} \label{table:params}
\end{table}

For numerical estimates of the sensitivity of the discussed sensor, we consider the fused silica microresonator with the parameters listed in Table \ref{table:params}, see Refs.\,\cite{96a1GoIlSa, 20a1BaKhStMaBi}.

\begin{figure}
  \includegraphics[width=0.45\textwidth]{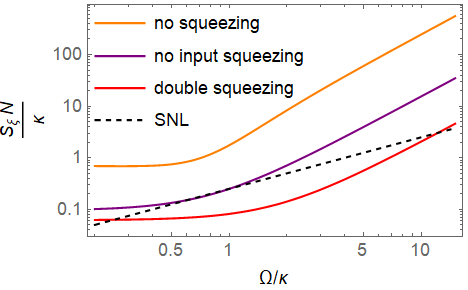}
  \caption{Plots of spectral densities \eqref{S_0}, \eqref{S_1}, \eqref{S_3}, and \eqref{S_SNL} for  the parameters in Table \ref{table:params}}\label{fig:plots}
\end{figure}

In Fig.\,\ref{fig:plots}, the spectral densities \eqref{S_0}, \eqref{S_1}, and \eqref{S_3}, together with the SNL \eqref{S_SNL}, normalized by the factor $\kappa/N$, are plotted assuming these values. These plots clearly show the sensitivity gain provided by the squeezing. In particular, these plots show that using the squeezing, the SNL can be overcome in a broad frequency band  that exceeds the microresonator bandwidth $\kappa$.

The main results of this paper can be summarized as follows. We calculated the sensitivity limitations of the optical microresonator-based sensors, imposed by quantum fluctuations of the probing light. We showed that using the squeezed quantum state of input light and additional intracavity squeezing of the light in the microresonator, it is possible to significantly increase the sensitivity, overcoming the SNL in a broad frequency band. The negative nonlinear effect - SPM - can also be suppressed by means of intracavity squeezing. In this case, the sensitivity will be limited only by the optical losses and the available degree of squeezing.

\acknowledgments

The work of D.S., I.B. and F.K. was supported by the Russian Science Foundation (Project No. 25-12-00263).

\appendix

\section{Output signal}\label{app:d_zeta}

The solution to Eqs.\,\eqref{b_cs} is the following:
\begin{subequations}
  \begin{multline}
    \hat{b}_c = \frac{\sqrt{2}}{\mathcal{D}}\bigl[
        k_s\beta\xi + (\ell+k_c)(\sqrt{\kappa'}\hat a_c + \sqrt{\kappa''}\hat v_c) \\
        - k_s(\sqrt{\kappa'}\hat a_s + \sqrt{\kappa''}\hat v_s)
      \bigr] ,
  \end{multline}
  \begin{multline}
    \hat{b}_s = \frac{\sqrt{2}}{\mathcal{D}}\bigl[
        -(\ell-k_c)\beta\xi
        - k_\gamma(\sqrt{\kappa'}\hat a_c + \sqrt{\kappa''}\hat v_c) \\
        + (\ell-k_c)(\sqrt{\kappa'}\hat a_s + \sqrt{\kappa''}\hat v_s
      \bigr] ,
  \end{multline}
\end{subequations}
where
\begin{equation}\label{D}
  \mathcal{D} = \ell^2 - k_c^2 - k_sk_\gamma \,.
\end{equation}
Substitution of this solution into Eq.\,\eqref{d_cs} and then into Eq.\,\eqref{d_zeta_raw}
gives that
\begin{multline}\label{d_zeta_app}
  \hat d_{\zeta} = \frac{2\sqrt{\eta\kappa'}}{D}\bigl\{
      [k_s\cos\zeta - (l-k_c)\sin\zeta]\beta\xi \\
      + [
          (\ell + k_c)(\sqrt{\kappa'}\hat a_c + \sqrt{\kappa''}\hat v_c)
          - k_s(\sqrt{\kappa'}\hat a_s + \sqrt{\kappa''}\hat v_s)
        ]\cos\zeta \\
      + [
          -k_\gamma(\sqrt{\kappa'}\hat a_c + \sqrt{\kappa''}\hat v_c)
          + (\ell - k_c)(\sqrt{\kappa'}\hat a_s + \sqrt{\kappa''}\hat v_s)
        ]\sin\zeta
      \bigr\} \\
    - \sqrt{\eta}\hat{a}_{\zeta} + \sqrt{1-\eta}\hat{u}_{\zeta} \\
  = \frac{\sqrt{\eta}}{\mathcal{D}}(G\xi + \hat{w})
  = \frac{\sqrt{\eta}G}{\mathcal{D}}(\xi + \hat{\xi}_{\rm fl})\,,
\end{multline}
where
\begin{equation}\label{xi_fl}
  \hat{\xi}_{\rm fl} = \frac{\hat{w}}{G}
\end{equation}
is the total noise, reduced to the meter input,
\begin{subequations}\label{G}
  \begin{equation}
    G = G_c\cos\zeta + G_s\sin\zeta \,,
  \end{equation}
  with
  \begin{equation}\label{Gcs}
    G_c = 2\sqrt{\kappa'}\beta k_s \,, \quad G_s = -2\sqrt{\kappa'}\beta(\ell - k_c) \,,
  \end{equation}
\end{subequations}
is the gain factor,
\begin{equation}\label{w}
  \hat{w} = A_c\hat{a}_c + A_s\hat{a}_s + V_c\hat{v}_c + V_s\hat{v}_s
    + \epsilon\mathcal{D}\hat{u}_\zeta \,,
\end{equation}
and
\begin{subequations}\label{A_1}
  \begin{gather}
    A_c = A_{cc} \cos\zeta + A_{cs}\sin\zeta \,, \quad
    A_s = A_{sc} \cos\zeta + A_{ss}\sin\zeta \,, \\
    V_c = V_{cc} \cos\zeta + V_{cs}\sin\zeta \,, \quad
    V_s = V_{sc} \cos\zeta + V_{ss}\sin\zeta \,,
  \end{gather}
\end{subequations}
\begin{subequations}\label{A_11}
  \begin{gather}
    A_{cc} = 2\kappa'(\ell+k_c) - \mathcal{D} \,, \quad
    A_{cs} = -2\kappa'k_\gamma \,, \\
    A_{sc} = -2\kappa'k_s \,, \quad
    A_{ss} = 2\kappa'(\ell-k_c) - \mathcal{D} \,, \\
    V_{cc} = 2\sqrt{\kappa'\kappa''}(\ell+k_c) \,, \quad
    V_{cs} = -2\sqrt{\kappa'\kappa''}k_\gamma \,, \\
    V_{sc} = -2\sqrt{\kappa'\kappa''}k_s \,, \quad
    V_{ss} = 2\sqrt{\kappa'\kappa''}(\ell-k_c) \,,
  \end{gather}
\end{subequations}

In the case of the squeezed input state, spectral densities of the quadratures $\hat{a}_{c,s}$ and their cross-correlation spectral density are equal to
\begin{subequations}\label{S_11}
  \begin{gather}
    S_{cc} = \tfrac12(\ch2r + \sh2r\cos2\theta) \,, \\
    S_{ss} = \tfrac12(\ch2r - \sh2r\cos2\theta) \,, \\
    S_{cs} = -\tfrac12\sh2r\sin2\theta \,,
  \end{gather}
\end{subequations}
while spectral densities of all other noise quadratures are equal to $1/2$. Therefore, spectral density of the noise \eqref{w} is equal to
\begin{multline}\label{S_w_gen}
  S_w = |A_c|^2S_{cc} + |A_s|^2S_{ss} + 2\Re(A_c^*A_s)S_{cs} \\
    + \tfrac12(|V_c|^2 + |V_s|^2 + \epsilon^2|D|^2) \,,
\end{multline}
and spectral density of the reduced to the input noise \eqref{xi_fl} is equal to
\begin{equation}\label{S_xi}
  S_\xi = \frac{S_w}{|G|^2} \,.
\end{equation}

\section{Optimal parameters}\label{app:opt_theta_zeta}

\subsection{Step 1}\label{app:opt_theta_zeta_1}

In this Appendix, we assume that
\begin{equation}\label{Omega_0}
  \Omega=0 \,.
\end{equation}
In this case,
\begin{equation}
  \ell = \kappa\,,\quad \mathcal{D} = \kappa^2 - k_c^2 - k_sk_\gamma
\end{equation}
and all numerical factors \eqref{A_11} are real. Correspondingly, spectral density \eqref{S_w_gen} takes the following form:
\begin{equation}
  S_w = \tfrac12[A_c^2S_{cc} + A_s^2S_{ss} + 2A_cA_sS_{cs} \\
    + V_c^2 + V_s^2 + \mathcal{D}^2\epsilon^2] \,.
\end{equation}

The minimum of this equation in $\theta$ corresponds to
\begin{equation}\label{opt_theta}
  \tan2\theta = \frac{2A_cA_s}{A_s^2 - A_c^2}
\end{equation}
and is equal to
\begin{multline}
  S_w = \tfrac12[(A_c^2 + A_s^2)e^{-2r} + V_c^2 + V_s^2 + \mathcal{D}^2\epsilon^2] \\
  = \tfrac12[W_c\cos^2\zeta + 2W_{cs}\cos\zeta\sin\zeta + W_s\sin^2\zeta] ,
\end{multline}
where
\begin{subequations}\label{W}
  \begin{gather}
    W_c = (A_{cc}^2 + A_{sc}^2)e^{-2r} + V_{cc}^2 + V_{sc}^2 + \epsilon^2\mathcal{D}^2\,, \\
    W_s = (A_{cs}^2 + A_{ss}^2)e^{-2r} + V_{cs}^2 + V_{ss}^2 + \epsilon^2\mathcal{D}^2\,, \\
    W_{cs} = (A_{cc}A_{cs} + A_{sc}A_{ss})e^{-2r} + V_{cc}V_{cs} + V_{sc}V_{ss} \,.
  \end{gather}
\end{subequations}

Therefore,
\begin{equation}
  S_\xi
  = \frac{1}{2}\frac{W_c + 2W_{cs}\tan\zeta + W_s\tan^2\zeta}{(G_c + G_s\tan\zeta)^2}  \,.
\end{equation}
The minimum of this spectral density in $\tan\zeta$ is given by
\begin{equation}\label{opt_zeta}
  \tan\zeta = \frac{G_sW_c - G_cW_{cs}}{G_cW_s - G_sW_{cs}}
\end{equation}
and is equal to
\begin{equation}\label{S_W}
  S_\xi = \frac{1}{2}\frac{W_cW_s - W_{cs}^2}{G_c^2W_s - 2G_cG_sW_{cs} + G_s^2W_c} \,.
\end{equation}

\subsection{Step 2}\label{app:opt_stage2}

The values of the deviation $k_s-2\gamma N$, numerically calculated for the optimized values of $k_s$ assuming the typical values of the parameters, are listed in Table \ref{tab:dk}.

\begin{table}[h]
\centering
\begin{ruledtabular}
\begin{tabular}{ccccc}
$N$ & $r$ & $\eta$ & $\kappa_1=0.9$ & $\kappa_1=0.7$ \\
\hline

\multirow{6}{*}{$10^6$}
& \multirow{2}{*}{0}
& 0.9 & -1.224 & -0.897 \\
& & 0.7 & -1.556 &  0.500 \\ \cline{2-5}

& \multirow{2}{*}{0.69}
& 0.9 &  0.541 & -0.477 \\
& & 0.7 & -1.344 & -2.555 \\ \cline{2-5}

& \multirow{2}{*}{1.7}
& 0.9 & -4.869 &  2.876 \\
& & 0.7 &  3.508 & -27.037 \\

\hline

\multirow{6}{*}{$10^7$}
& \multirow{2}{*}{0}
& 0.9 & -0.104 & -0.518 \\
& & 0.7 & -1.151 &  0.467 \\ \cline{2-5}

& \multirow{2}{*}{0.69}
& 0.9 & -0.217 & -1.959 \\
& & 0.7 &  1.805 &  1.155 \\ \cline{2-5}

& \multirow{2}{*}{1.7}
& 0.9 & -0.758 & -7.964 \\
& & 0.7 & -11.552 & -17.788 \\

\hline

\multirow{6}{*}{$10^8$}
& \multirow{2}{*}{0}
& 0.9 &  1.154 & -0.811 \\
& & 0.7 &  0.039 &  0.430 \\ \cline{2-5}

& \multirow{2}{*}{0.69}
& 0.9 &  0.554 & -0.864 \\
& & 0.7 &  0.074 &  1.199 \\ \cline{2-5}

& \multirow{2}{*}{1.7}
& 0.9 & -3.823 & -1.402 \\
& & 0.7 & -6.866 & -13.951 \\

\end{tabular}
\end{ruledtabular}
\caption{The values of $(k_s-2\gamma N)\times10^8$.}
\label{tab:dk}
\end{table}

\subsection{Step 3}\label{app:opt_theta_zeta_3}

In the case of conditions \eqref{k_s_opt} and \eqref{Omega_0}, Eqs.\,\eqref{A_11} take the following form:
\begin{subequations}\label{A_11_noSFM}
  \begin{gather}
    A_{cc} = (\kappa+k_c)(\kappa'-\kappa''-k_c) \,, \quad  A_{cs} = 0 \,, \\
    A_{sc} = -2\kappa'k_s \,, \quad A_{ss} = 2(\kappa-k_c)(\kappa'-\kappa''-k_c) \,, \\
    V_{cc} = 2\sqrt{\kappa'\kappa''}(\kappa+k_c) \,, \quad  V_{cs} = 0 \,, \\
    V_{sc} = -2\sqrt{\kappa'\kappa''}k_s \,, \quad
    V_{ss} = 2\sqrt{\kappa'\kappa''}(\kappa-k_c) \,.
  \end{gather}
\end{subequations}
Substitution of these values into the denominator of Eq.\,\eqref{opt_zeta} gives that
\begin{multline}
  G_cW_s - G_sW_{cs} = 2\sqrt{\kappa'}\beta k_s(\kappa-k_c)(\kappa+k_c) \\
    \times[k_c(e^{-2r} + \epsilon^2) - (\kappa' - \kappa'')e^{-2r} + \epsilon^2\kappa] \,.
\end{multline}
It is easy to see that in the case of \eqref{k_c_opt}, this value is equal to zero, which gives \eqref{zeta_opt}. In this case, it follows from  Eq.\,\eqref{A_11} that $A_c=0$, which, taking into account Eq.\,\eqref{opt_theta}, gives  \eqref{theta_opt}.

\section{No squeezing case}\label{app:no_sqz}

In the case of Eq.\,\eqref{no_sqz}, Eqs.\,\eqref{D}, \eqref{Gcs} and \eqref{A_11} take the following form:
\begin{gather}
  \mathcal{D} = \ell^2 \,, \\
  G_c = 0 \,, \quad G_s = -2\sqrt{\kappa'}\beta\ell \,, \label{Gcs0}
\end{gather}
\begin{subequations}\label{A_11_0}
  \begin{gather}
    A_{cc} = A_{ss} = \ell(2\kappa' - \ell) \,, \\
    A_{cs} = -2\kappa'k_\gamma \,, \quad  A_{sc} = 0 \,, \\
    V_{cc} = V_{ss} = 2\sqrt{\kappa'\kappa''}\ell \,, \\
    V_{cs} = -2\sqrt{\kappa'\kappa''}k_\gamma \,,\quad V_{sc} = 0 \,.
  \end{gather}
\end{subequations}
Therefore, spectral density of the noise \eqref{w} is equal to
\begin{multline}
  S_w(\Omega) = \tfrac12(|A_c|^2 + |A_s|^2 + |V_c|^2 + |V_s|^2 + \epsilon^2|\mathcal{D}|^2)
    \\
  = \tfrac12[
        |A_{cc}|^2 + |V_{cc}|^2
        + 2\Re{(A_{cc}A_{cs}^* + V_{cc}V_{cs}^*)}\cos\zeta\sin\zeta \\
        + (|A_{cs}|^2 + |V_{cs}|^2)\sin^2\zeta + \epsilon^2|\mathcal{D}|^2
      ] .
\end{multline}
The corresponding spectral density of the normalized noise $\xi_{\rm fl}$ is equal to
\begin{equation}\label{S_xi_0}
  S_\xi(\Omega) = \frac{S_w(\Omega)}{|G_s|^2\sin^2\zeta}
  = \frac{1}{8N}\biggl(
        \frac{|\ell|^2}{\eta\kappa'\sin^2\zeta} - 4k_\gamma\cot\zeta
        + \frac{4\kappa k_\gamma^2}{|\ell|^2}
      \biggr) .
\end{equation}

At $\Omega=0$, the minimum of this equation is achieved at
\begin{equation}
  \cot\zeta = \frac{2\eta\kappa'k_\gamma}{\kappa^2} \,.
\end{equation}
Substitution of this value into Eq.\,\eqref{S_xi_0} gives Eq.\,\eqref{S_0}.

\providecommand{\noopsort}[1]{}\providecommand{\singleletter}[1]{#1}%

\end{document}